\documentclass[aps,prb,twocolumn,groupedaddress,showpacs,letterpaper]{revtex4}

\usepackage[dvips]{graphics}

\begin{document}
\title{Observation of mixed anisotropy in the critical susceptibility of an ultrathin magnetic film}
\author{K. Fritsch} \author{R. D'Ortenzio} \author{D. Venus}
\email{venus@physics.mcmaster.ca} 
\affiliation{Department of Physics
and Astronomy, McMaster University, Hamilton, Canada}
\date{\today}

\begin{abstract}

Measurements of the magnetic susceptibility of ultrathin Fe/W(110) films with thickness in the range of 1.6 to 2.4 ML Fe, show that in addition to the large response along the easy axis associated with the Curie transition, there is a much smaller, paramagnetic hard axis response that is not consistent with the 2D anisotropic Heisenberg model used to describe homogeneous in-plane ferromagnets with uniaxial anisotropy.  The shape, amplitude, and peak temperature of the hard axis susceptibility, as well as its dependence upon layer completion close to 2.0 ML, indicate that inhomogeneities in the films create a system of mixed anisotropy.  A likely candidate for inhomogeneities that are magnetically relevant in the critical region are the closed lines of step edges associated with incomplete layers.  According to the Harris criterion, the existence of magnetically relevant inhomogeneities may alter the critical properties of the films from those of a 2D Ising model.  Experiments in the recent literature are discussed in this context.

\end{abstract}
\maketitle

\section{Introduction}

The development of the ability to grow and experimentally investigate ultrathin ferromagnetic films over the last few decades has provided an opportunity to address ongoing fundamental questions regarding phase transitions in two dimensional (2D) systems from a fresh perspective.  One such enduring question is the role of magnetic inhomogeneities or defects in altering the behavior of the phase transitions of a 2D magnetic system.  This is important because essentially all physically realizable materials have defects.  The present article explores these issues using magnetic susceptibility measurements of the critical region of Fe/W(110) ultrathin films with between 1.6 and 2.4 Fe ML. 

Ideal ferromagnetic 2D films with in-plane uniaxial anisotropy are predicted to display a range of behaviors, depending upon the proximity of the temperature to the Curie temperature $T_C$.  These films are described by a 2D anisotropic Heisenberg Hamiltonian, where the crystalline and surface anisotropies are most often modelled as an exchange interaction that is anisotropic at the level of 1\% to 0.1\%.  Both theory\cite{bander} and simulations\cite{serena} indicate that the presence of this small anisotropy moves the Curie temperature from 0 K to of order 100s of K, and induces the critical behavior of a 2D Ising model.  A growing number of experimental measurements of the critical exponent of the magnetization, $\beta$, confirm this.\cite{taroni}  Simulations\cite{cuccoli} indicate that the cross-over from 2D anisotropic Heisenberg to 2D Ising behavior occurs relatively far from $T_C$, and, to our knowledge, has not yet been observed experimentally.  The hard axis susceptibility in the critical regime can also be derived from the 2D anisotropic Heisenberg model.\cite{jensen}  It is a measure of the microscopic anisotropy that persists in the paramagnetic region, and is described qualitatively by the 2D Ising model.\cite{fischer}

Real films have structural inhomogeneities that can easily create local magnetic inhomogeneities.  Examples are missing atoms (site dilution), structural defects or atomic adsorption that can locally alter the  exchange coupling or the microscopic anisotropy, or both. The specific example of a sample where sites have one of two different anisotropies is termed ``mixed anisotropy''.\cite{remark}  Central questions are whether these inhomogeneities are relevant or irrelevant to the critical behavior, and how they affect the extent of the asymptotic region.  The Harris criterion\cite{harris} treats the case of random, point-like inhomogeneities. Experiments on 3D systems have confirmed the Harris criterion.\cite{chan,birgeneau,birgeneau2}  In particular, for 3D Ising systems, introducing inhomogeneities in the form of either altered exchange coupling along the Ising axis\cite{birgeneau}, or local XY anisotropy perpendicular to the Ising axis\cite{birgeneau2} gives altered critical exponents in agreement with the 3D site diluted Ising model.\cite{jug2}  In two dimensions, inhomogeneities become relatively more important.  In the case of the 2D Ising model, the Harris criterion is not predictive.  This has lead to a large body of theoretical and numerical studies of the effect of inhomogeneities in this system.\cite{kenna}

If the inhomogeneities are not point-like and random, but correlated to form a line, then it may be that the effective dimensionality of the system is changed,\cite{igloi} so that the prediction of the Harris criterion is altered.\cite{harris,mccoy,bariev,mccoy2}

The current study investigates Fe/W(110) ultrathin films near the Curie transition, using measurements of the magnetic susceptibility.  This system has very strong uniaxial, in-plane anisotropy\cite{fritzsche} and is a model for 2D Ising behavior.  An experimental study of the magnetization in the ferromagnetic phase by Back \emph{et al.}\cite{back} has shown 2D Ising power law scaling over 18 orders of magnitude in reduced variables. In addition to the divergent, easy axis response, the present work reports a small response in the hard axis, in-plane magnetic susceptibility in the paramagnetic phase.  The form of the  hard axis response and its systematic dependence on the film thickness and layer completion demonstrates that it is due to inhomogeneities that create a system with mixed anisotropy. A likely candidate for the relevant inhomogeneities is the closed lines of atomic steps that are present when the film thickness deviates from two complete monolayers.  The implications of these magnetic inhomogeneities for the critical behavior is discussed with reference to recent experimental findings of the variation of the critical exponent of the susceptibility, $\gamma$, with layer completion in Fe/W(110).\cite{dunlavy}

\section{Experimental Results}

Ultrathin Fe films grown on W(110) in ultrahigh vacuum have been studied intensively, and have revealed magnetic behavior that is sensitive to film thickness, structure and cleanliness.  The first monolayer of iron wets the W(110) surface very well\cite{elmers1} and exhibits an in-plane crystalline anisotropy along [1-10].\cite{fritzsche}  The interfacial anisotropy of 1.9 mJ/m$^2$ is one of the largest known for an ultrathin film system.  According to scanning tunneling microscopy (STM) images, the structure of the second monolayer depends upon the substrate orientation and temperature during growth.  Deposition at or near room temperature on a substrate cut at an angle $\le 0.3^o$ to the [110] surface normal produces islands of 2ML thickness\cite{durkop,back,weber,kubetzka} which grow and begin to coalesce near a thickness of 1.6 ML. There is good layer completion near 2.0 ML, with small regions of 1 ML and 3 ML Fe coexisting.  If the substrate is miscut at an angle $\ge 0.5^o$, and the Fe is deposited near 660 to 700K, then the presence of many atomic steps and the increased mobility of the Fe atoms induces step-flow growth,\cite{elmers1,elmers3,hauschild} resulting in long ``strips'' of 2 ML thickness parallel to the substrate steps, alternating with strips of 1 ML or 3 ML.

In the present experiments, the crystal miscut is $\le 0.2^o$ as determined using \emph{ex situ} STM images.\cite{dunlavy2} The crystal was cleaned according to standard procedures of alternate heating in $1 \times 10^{-7}$ Torr oxygen, and flashing to white heat until the surface carbon contamination was not detectable using Auger electron spectroscopy, and a sharp W(110) LEED pattern was obtained.  The first monolayer was deposited at room temperature and annealed for 150 s at 500 K.  A second deposition of between 0.6 and 1.4 ML to complete the film was made at room temperature. LEED images confirmed that the Fe films were pseudomorphic with the substrate in this range of coverages\cite{gradmann}.  Although \emph{in situ} STM imaging was not possible, layer growth through the coalescence of monolayer islands is expected based upon the previously cited studies.  In order to confirm this expectation, the film growth was monitored using Auger electron spectroscopy and pseudomorphic growth was confirmed using low energy electron diffraction.  Fig. 1a presents the attentuation of the W(110) Auger signal as Fe was evaporated onto it in a series of sequential depositions.  After each deposition, the film was annealled to 700K.  The clear break in the plot illustrates the completion of one monolayer, where the W(110) Auger signal is attenuated to $0.60\pm0.02$ of the value for a clean substrate.
\begin{figure}
\vspace{-.25 in} 
\scalebox{.4}{\includegraphics{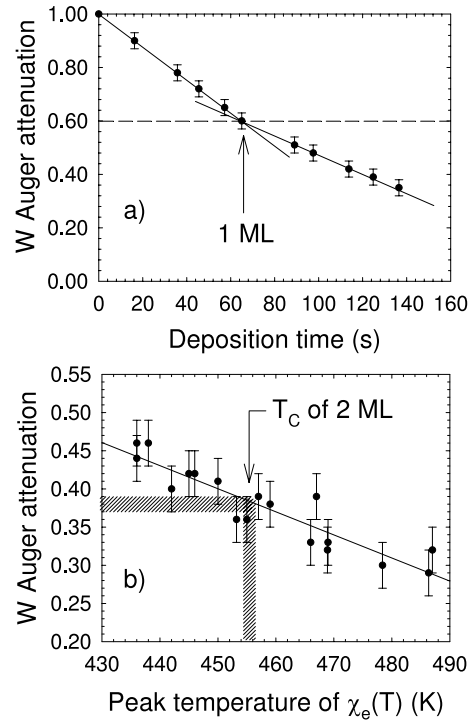}} 
\vspace{-.50in} 
\caption{a) The film thickness was calibrated by plotting the attenuation of the W(110) substrate signal of a Fe film grown in sequential steps against the deposition time.  The film was annealed at each step.  The break in the curve indicates a thickness of 1 ML and good wetting of the substrate.  b) The W Auger attenuation is plotted against Curie temperature (peak in $\chi_{[1-10]}$) for the films in this study.  The literature value of the Curie temperature for 2 ML Fe/W(110) corresponds to an attenuation of 0.38$\pm$0.01, in very good agreement with the expected value for layer-by-layer growth of 2 ML, 0.36$\pm$0.024.}
\end{figure}

Fig. 1b plots the Auger attenuation coefficient of the W substrate upon deposition of the entire Fe film, as a function of the Curie temperature, for the films used in this study.  Layer growth of a second complete monolayer is expected to correspond to an attenuation of $(0.60\pm0.02)^2 = 0.36\pm0.024$. According to the literature,\cite{elmers4,dunlavy} the Curie temperature of 2 ML Fe/W(110) is $455\pm3$K. In fig. 1b, the W Auger signal is in fact attenuated by 0.38$\pm$0.01 at this temperature, independantly confirming very good layer completion at 2 ML.  The fact that the W signal is not attenutated quite as much as expected indicates a residual small net area of 1 and 3 ML regions. The magnetic properties, as indicated by $T_C$, are consistent with previous studies where island growth has been demonstrated by STM.\cite{elmers4}  Since the Curie temperature is a much more sensitive measure of relative thickness than the Auger attenuation, Fe thicknesses quoted in the remainder of this article are calculated based on the Curie temperature and the calibration curves in fig. 1.

\begin{figure}
\vspace{-.25 in} 
\scalebox{.4}{\includegraphics{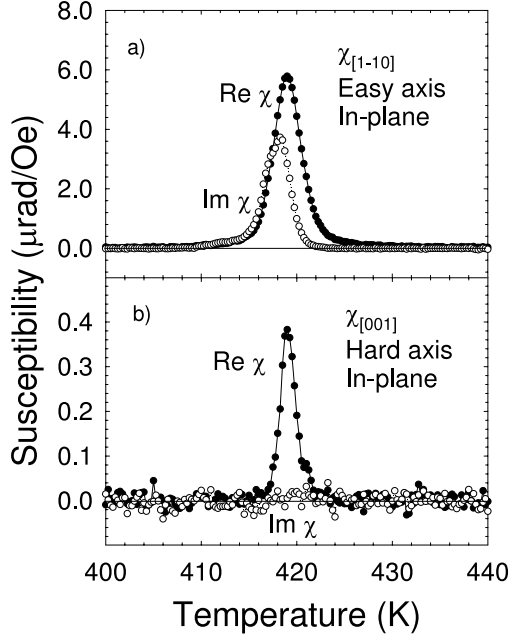}} 
\vspace{-.50in} 
\caption{Real (solid dots) and imaginary (open dots) parts of the magnetic susceptibility of 1.7 ML Fe/W(110).  a)$\chi_{[1-10]}(T)$ b) $\chi_{[001]}(T)$ }
\end{figure}
Previous studies\cite{elmers3} have shown that isolated 2 ML Fe islands on 1 ML Fe/W(110) have perpendicular magnetic anisotropy at temperatures below about 235 K, but that once the islands coalesce at a thickness of about 1.6 ML, there is once again strong in-plane, uniaxial anisotropy.  At lower temperatures, the perpendicular anisotropy persists to higher coverage.\cite{kubetzka,kubetzka2}  The perpendicular anisotropy of the islands is very sensitive to gas adsorption,\cite{durkop} and a small gas exposure induces a spin-reorientation transition that leaves the entire sample with in-plane anisotropy.   Since the films in the present study were all of thicknesses greater than 1.6 ML Fe, and the temperature range investigated was 400 - 480 K, only in-plane anisotropy was observed, consistent with these results.

Measurements of the magnetic susceptibility were made \emph{in situ} using the surface magneto-optic Kerr effect with an applied ac field of 150 Hz and lock-in amplification.\cite{arnold}  Recent modifications of the apparatus\cite{fritsch} that increased structural rigidity, reduced scattered light from the polarizing crystals, and dispensed with apertures, have increased the sensitivity and reduced the noise so that optical rotations of 15 nrad are detectable.  The measurements use a pair of field coils aligned very nearly parallel to the sample surface, and in the scattering plane of the laser light.  The light is incident at $45^o$ to the sample surface, so that the measured ellipticity is a combination of the longitudinal and polar Kerr effects.  The sample can be rotated about its normal, so that the aligned scattering plane and applied field can lie along any in-plane direction.  For definiteness, the measured susceptibilities are labelled with a subscript that denotes this in-plane direction.  The susceptibility is given in units of $\mu$radians/Oe, since these can be calibrated absolutely.

Figure 2 presents measurements of the magnetic susceptibility for a 1.7 ML Fe/W(110) film using an ac field of 0.50 Oe. Part a) contains the real and imaginary parts of $\chi_{[1-10]}(T)$.  The narrow peaks in both the real and imaginary easy axis response are characteristic of a Curie transition and the associated hysteretic dissipation just below $T_C$.  This result is entirely as expected for remanent magnetization along the easy axis.  Part b) illustrates the hard axis susceptibility $\chi_{[001]}(T)$. This curve is different from the easy axis result in three important ways: the amplitude is smaller by a factor of 15, the full width at half maximum (FWHM) of the peak is significantly narrower, and there is no measurable imaginary response. 

A systematic study of this film as a function of the crystal orientation is presented in figure 3.  
\begin{figure}
\vspace{-.25 in} 
\scalebox{.4}{\includegraphics{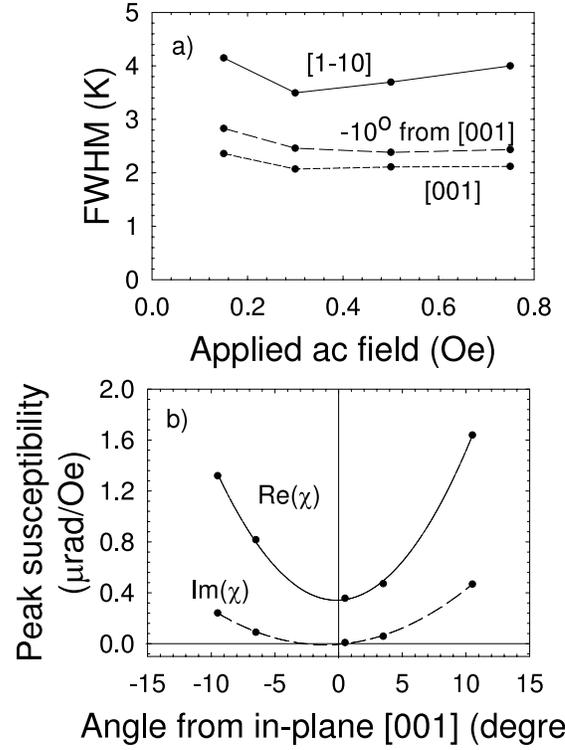}} 
\vspace{-.40in} 
\caption{a) The FWHM of the susceptibility peak is plotted as a function of the applied ac field amplitude, for three orientations of the scattering plane and field direction; [1-10], [001] and between these orientations, $10^o$ from the [001].  b) The maximum amplitude of the real and imaginary parts of the susceptibility are plotted as a function of the angle that both the scattering plane and direction of applied field make with the [001] hard axis.}
\end{figure}
Part a) presents the FWHM of the real part of the susceptibility as a function of the applied ac field measured along the easy [1-10] axis, the hard [001] axis, and an intermediate in-plane angle $10^o$ from the hard axis.  These results make it clear that the peak widths are not due to finite field effects, and quantifies the fact that the hard axis response is narrower than the easy axis response by nearly a factor of 2.   Fig. 3b presents the amplitude of the real and imaginary parts of the susceptibility as a function of the angle $\phi$ that the in-plane hard axis makes with the scattering plane and applied field.  Both of these amplitudes are well described by a quadratic variation about the hard axis, where the imaginary response goes to zero.  This confirms that the measured hard axis response is a distinct signal that has no imaginary part.  Moving away from the hard axis creates a component of the applied field along the easy axis, and the measured susceptibility is a mixture of the easy and hard axis responses that is dominated by the much larger easy axis susceptibility.

All of the above experiments were performed with p-polarized light. Measurements with s-polarized light (not shown) produced similar susceptibility traces.  This excludes the possibility that the signals labeled $\chi_{[001]}$ arise through the transverse Kerr effect with the magnetization perpendicular to the scattering plane.\cite{remark2}  Measurements were also made with a separate magnetic coil aligned very nearly with the surface normal hard axis, [110] (not shown).  These signals were a further order of magnitude smaller than those along the in-plane hard axis, and had the shape of the easy axis signal when the scattering plane included the [1-10] and [110] direction, and the shape of the in-plane hard axis signal when the scattering plane included the [001] and [110] directions.\cite{fritsch}  It was therefore concluded that these are not true measurements of the susceptibility normal to the surface, but rather small components of the in-plane susceptibilities due to a small misalignment of the coil and crystal axes that could not be removed using the degrees of freedom of the crystal holder.

The results of fig. 2 and 3 indicate that these films have a reproducible, narrow hard axis response that contains no imaginary component - i.e. it is dissipationless.  In order to better characterize the system, susceptibility measurements have been made on a collection of 22 separate films with thicknesses in the range of 2.0 $\pm$0.4 ML of Fe.  The main panel of figure 4a presents three curves measured sequentially from a 2.2 ML film.  First, the real part of the easy axis susceptibility (solid circles) was 
\begin{figure}
\vspace{-.25 in} 
\scalebox{.4}{\includegraphics{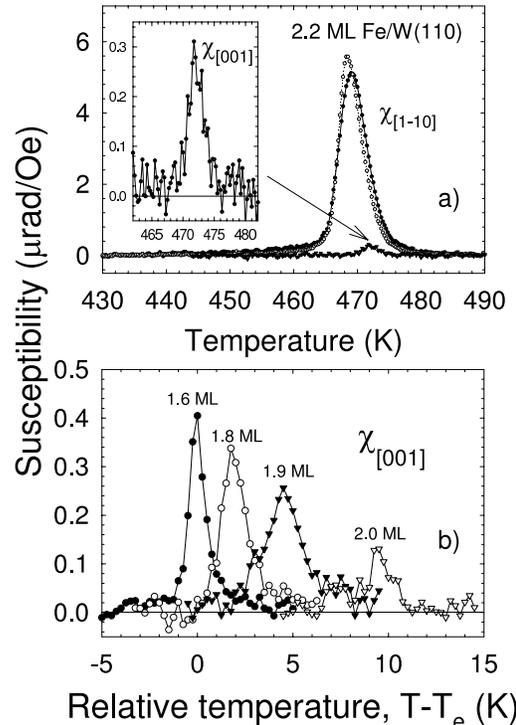}} 
\vspace{-.65in} 
\caption{a) The real part of the susceptibility of a 2.2 Fe ML film measured three times in sequence:  along the easy axis (solid circles), the hard axis (solid triangles, and inset), and along the easy axis again (open circles).  The hard axis signal peaks about 3 K above the easy axis signal.  The latter is reproduced to within 1 K.  b) The hard axis [001] susceptibility is plotted for films of different Fe thickness, with a temperature scale measured relative to the peak temperature of the corresponding easy axis [1-10] susceptibility.  The uncertainty in this relative scale is $\pm$ 1K.}
\end{figure}
measured, then the sample was rotated and the hard axis susceptibility (solid triangles, and in more detail in the inset) was measured.  The sample was then rotated back to its original alignment and the easy axis susceptibility was remeasured (open circles).  For this film the peak of the hard axis susceptibility occurs about 3K higher in temperature than the peak of the easy axis susceptibility.  This is not because of thermally induced changes in the film, since the easy axis signal is reproducible to within 1 K after measuring the hard axis curve. This reproducibility was checked on 1/3 of the films, either by sequential easy-hard-easy axis measurements or by repeated measurements of the easy or hard axis susceptibilities. This includes samples showing hard axis susceptibility peaks up to 10 K above $T_C$.  An uncertainty of $\pm 1$K of the peak positions is a reliable representation of our measurements, as no instance of a larger change in $T_C$ upon re-measurment was observed.

Fig. 4b shows hard axis [001] susceptibility of a selection of films, labelled by their thickness.  Note that the temperature scale is \emph{relative} to the peak of the easy axis susceptibility for each film, $T_e$, which closely tracks $T_C$.  The uncertainty in this relative scale is $\pm$1 K, as discussed above.  These curves illustrate that the hard axis response depends upon the film thickness in two important ways:  first, the amplitude of the susceptibility decreases as the film thickness moves closer to 2.0 ML;  second, there is a correlated  shift in the peak temperature, so that the hard axis peak moves further into the paramagnetic phase as the film approaches a thickness of 2.0 ML.  In all cases, there is no imaginary response.  

Figure 5 summarizes the results for the entire collection of films.  Part a) plots 
\begin{figure}
\vspace{-.25 in} 
\scalebox{.4}{\includegraphics{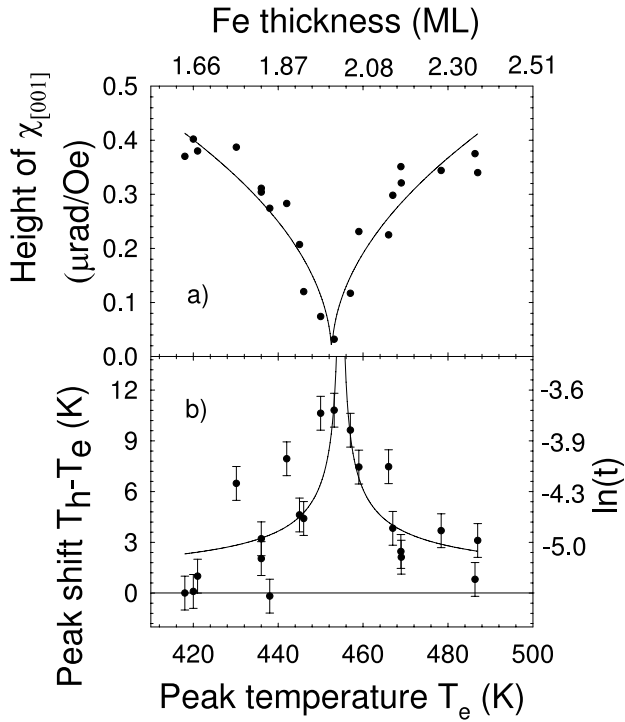}} 
\vspace{-.55in} 
\caption{Features of the hard axis susceptbility for 25 different films are summarized.  a) The peak amplitude is plotted against $T_e$.  The thickness scale on the top of the plot is derived using the data in fig. 1. b) The peak position, relative to the peak temperature of the easy axis susceptibility of the same film.  The non-linear right hand scale uses the reduced temperatue $t=T_h/T_e -1$.  The fitted curves are discussed in the text.}
\end{figure}
the amplitude of the hard axis susceptibility as a function of $T_e$.  Part b) plots the relative peak position of the hard axis response, $T_h$, compared to that of the easy axis susceptibility, $T_e$.  The error bars indicate the characteristic reproducibility upon repeated measurement, as was discussed previously.  These data confirm that the hard axis susceptibility is part of the paramagnetic, non-hysteretic response of the system as the incipient ferromagnetic state develops with the approach to $T_C$ from above.   Both the amplitude and the peak temperature of the paramagnetic hard axis response are strongly correlated with Fe coverage, $\theta$, as some function of $|\theta - \theta_0|$, where $\theta_0$= 2.0 ML.  The fitted lines are discussed in the next section.  The symmetry of the data about a thickness very close to 2 ML indicates that regions of 1 ML and 3 ML in a nominally 2 ML film produce equivalent effects in the hard axis susceptibility.

\section{Discussion}

The form of the hard axis susceptibility and its dependence upon $|\theta - \theta_0|$ indicates that these films are not homogeneous, anisotropic 2D Heisenberg systems.  Simulations,\cite{serena,cuccoli} theory and experiment\cite{jensen} indicate that in this model the hard axis susceptibility exhibits a very small, rounded, non-divergent contribution that is maximized at $T_C$ and persists below the Curie temperature, in qualitative agreement with the 2D Ising model.\cite{fischer}  Strongly anisotropic behavior persists relatively far from $T_C$.\cite{cuccoli}  In contrast, the present experiments observe a hard axis susceptibility that is sharply peaked in the paramagnetic region above $T_C$, does not persist down to $T_C$, and is only 1 to 2 orders of magnitude smaller than the easy axis signal.  In addition, the susceptibility of a system that is effectively uniform should not vary as a strong function of $|\theta - \theta_0|$.

The measurements are consistent instead with a system of mixed anisotropy.  Magnetic inhomogeneities that can produce mixed anisotropy arise naturally from the structural inhomogeneities implicit in films that are not complete monolayers.  In a system such as Fe/W(110), where layer-by-layer growth near 2 ML proceeds by the colescence of islands, the distribution of 1 ML pits and 3 ML islands will vary in number, perimeter and area as some function of $|\theta - \theta_0|$.  Changes in the magnitude and/or direction of anisotropy due to monolayer step edges,\cite{albrecht} atomic adsorption at step edges,\cite{durkop} or a monolayer change in thickness,\cite{elmers3} are well known.  All of these effects can give rise to a minority of the sample which has a local easy axis which is not aligned with the global easy axis of the 2D anisotropic Heisenberg system. The local easy axis response from these regions would then appear in the global [001] hard axis susceptibility.  

The functional form of the dependence of the paramagnetic hard axis susceptibility in fig. 5 can be understood qualitatively in terms of inhomogeneities of characteristic size $d$, by considering the changing correlation length, $\xi$, of the system as the temperature approaches $T_C$.   At high temperatures, the correlation length will be small, such that $\xi << d$.  In this case there will be essentially no coherent magnetic response from different regions of the inhomogeneity, and the hard axis susceptibility will be near zero.  At some  temperature closer to $T_C$, the condition $\xi \approx d$ will be met.  The entire inhomogeneity can then respond coherently and produce a maximum in the hard axis susceptibility. Even closer to $T_C,\; \xi >> d$.  In this circumstance, the inhomogeneity is immersed in a large correlated region with a uniaxial easy axis.  The resulting exchange stiffness within the correlated region inhibits the hard axis response of the defect atoms, and the hard axis susceptibility falls quickly once more.  This creates a peak in the hard axis susceptibililty. 

According to this qualitative picture, the temperature of the peak of the hard axis susceptibility, $T_h$, will occur when $\xi \approx d$.  Since the coverage $\theta$ represents a surface area on the film, the characteristic size or diameter of the inhomogeneity will scale as $d \sim \sqrt{|\theta - \theta_0|}$, giving
\begin{equation}
\frac{T_h -T_e}{T_e} \approx \frac{T_h -T_C}{T_C} \sim (\frac{\xi_0}{\xi(T_h)})^\nu \sim \frac{1}{d} \sim \frac{A}{\sqrt{|\theta-\theta_0|}},
\end{equation}
since $\nu =1$ for the 2D Ising model.  At temperature $T_h$, the maximum value of the hard axis susceptibility will scale as the total number of atoms in a characteristic inhomogeneity. This leads to scaling as $d^n$, where the (integer) $n$ depends upon the geometry of the inhomogeneity.
\begin{equation}
\label{amp}
\chi(T_h) \sim d^n \sim B(\sqrt{|\theta-\theta_0|})^n.
\end{equation}
In these expressions, $A$ and $B$ are constants.  For the fitted lines in fig. 5, the primary, more precise variable $\sqrt{|T_e - T_e^0|}$ has been used instead of the derived variable $\sqrt{|\theta-\theta_0|}$.  The fitted values for $T_e^0$ are 453.0$\pm$0.7 K and 454.7$\pm$0.5 K, in fig. 5a and 5b respectively.  Both values are very close to the Curie temperature of 2 ML Fe/W(110).  The best fit to the data in fig. 5a is given with $n=1$.

It is not possible to unambiguously identify the inhomogeneities responsible for the hard axis susceptibility in the present experiments, but the most likely candidate is the monolayer steps at the edges of regions of 1 or 3 ML Fe thickness.  Near a coverage of 2 ML, these step edges will form closed curves along the perimeter of 3 ML islands or 1 ML pits of characteristic size $d$ that will contribute to the susceptibility as $\sim d^1$. This agrees with the observation of identical magnetic effects for a local monolayer increase or decrease in thickness from 2 ML, since the creation of an island or pit gives an equivalent perimeter.  A less likely candidate is the areas of the 3 ML islands and 1 ML pits.  These do not normally have equivalent magnetic effects, as can be seen, for example, in the monotonic change in the Curie temperature with thickness.  In addition, areas scale as $n=2$ in eq.(\ref{amp}).  This gives a linear fit to a small range of data close to 2.0 ML, but does not represent all the data nearly as well.  It is unlikely that the observed changes in the anisotropy are due to adsorption, since then they would not depend upon layer completion, as observed, but rather the elapsed time.

Although it is obvious that there must be disorder at some level in a real system, these results demonstrate that the disorder related to layer completion is magnetically relevant and creates measurable, quantifiable effects in the critical region. Since the inhomogeneities create an additional response along a different axis than the bulk of the film, the disorder creates a film with mixed anisotropy.  This highlights the sensitivity of the paramagnetic susceptibility to inhomogeneities, since these measurements permit an observation of their explicit magnetic response.

\section{Implications for critical behavior}

These results are strong evidence that the Fe/W(110) films have mixed anisotropy, with the degree of disorder depending sensitively on the degree of layer completion near 2 ML.  This being the case, there are very general arguments concerning how these magnetic inhomogeneities will affect the critical properties of the films, that apply even though the nature of the magnetic inhomogeneities is not known precisely.  There are three inter-related aspects to consider.  The first is the local change in the magnitude of the anisotropy at an inhomogeneity and the resulting local variation in the Curie temperature.  The second is the characteristic size of the inhomogeneities, and the third is the overall degree of disorder created by the inhomogeneities.

Associated with the distribution of anisotropy directions and magnitudes produced by the inhomogeneities, there will also be a distribution of local Curie temperatures.
This is because the the Curie temperature of the anisotropic 2D Heisenberg model is determined by the anisotropy, and would be zero in the absence of an anisotropy.\cite{bander,serena}  An estimate of the Curie temperature for an anisotropic 2D Heisenberg system with exchange coupling $J$ and anisotropy $K$ is given by Serena \emph{et al.}\cite{serena}, as
\begin{equation}
T_C \approx \frac{T_C^I}{1+2\frac{T_C^I}{T_C^{SW}}}.
\end{equation}
In this expression, $T_C^I$ is the Curie temperature of the 2D Ising model with exchange coupling $J$, and $T_C^{SW}$ is the Curie temperature predicted by a first order spin wave treatment of the anisotropic 2D Heisenberg model.  The latter is given by
\begin{equation}
k_B T_C^{SW}=\frac{4\pi J}{\ln (J/K)}.
\end{equation}
For a small change in the anisotropy, the change in the Curie temperature is
\begin{equation}
\label{deltaTc}
\frac{\Delta T_C}{T_C} \approx \frac{1}{(2\pi/2.27)+\ln(J/K)} \frac{\Delta K}{K}.
\end{equation}
In the current system with an anisotropy of $K/J \approx 0.001$ and $T_C=450K$, a change in the Curie temperature of about 5K can be produced by a 10\% change in the anisotropy.

The Harris criterion\cite{harris} considers the question of whether or not these local variations in the Curie temperature will change the critical properties, for instance the critical exponents, of a system with inhomogeneities.  If the inhomogeneities are not point-like, but correlated, then the Harris criterion finds them relevant and able to change the critical properties.\cite{harris,mccoy,mccoy2}  The characteristic size of the inhomogeneity is important, since, in practical terms, an inhomogeneity of size $d$ is point-like if $\xi >> d$.  If $\xi \leq d$, then the inhomogeneity represents an independent region that contributes to the measured susceptibility.  The present experiments observe this independent contribution in the hard axis [001] susceptibility, where it is prominent because the contribution from regions with [1-10] easy axis anisotropy is so small.  The existence of this signal is an indicator that $\xi \leq d$, that the homogeneities are not point-like in this temperature range, and that critical exponents determined in this range are not likely to reflect universal properties.  The scaling of the peak temperature and amplitude with $\sqrt{\theta - \theta_0}$ indicates an increase in $d$ with layer disorder and a reduced range close to $T_C$ where the inhomogeneities can be considered point-like.

Finally, even if the inhomogeneities are point-like and randomly distributed, the overall degree of disorder can still be important.  The Harris criterion states that the relevance of inhomogeneities depends upon the sign of the critical exponent of the specific heat, $\alpha$.  Since the 2D Ising model has $\alpha =0$, the Harris criterion is not predictive.  However, theoretical\cite{dotsenko,shalaev,shankar,jug} and computational\cite{kenna,roder} work finds that the critical exponents are unchanged, but that additional logarithmic terms become important when the reduced temperature $t=T/T_C -1 < t_{cr}$, where $\ln t_{cr}=-1/g$ depends upon the overall degree of disorder through the parameter $g$.  The parameter $g \rightarrow 0$ as disorder disappears, but is not otherwise absolutely quantified except through numerical studies.  Numerical work by Roder {et al.}\cite{roder} has indicated that  fitting to a power law when logarithmic corrections are important produces
\begin{equation}
\label{geff}
\gamma_{eff} \approx \gamma_{Ising}[1+\frac{1}{2}\frac{\ln(|\ln t |)}{\ln \frac{1}{t}}].
\end{equation}
In this case $\gamma_{eff}$ is always greater than the 2D Ising value. 

In this context, the present experimental results bring new insight to a recent \begin{figure}
\vspace{-.25 in} 
\scalebox{.4}{\includegraphics{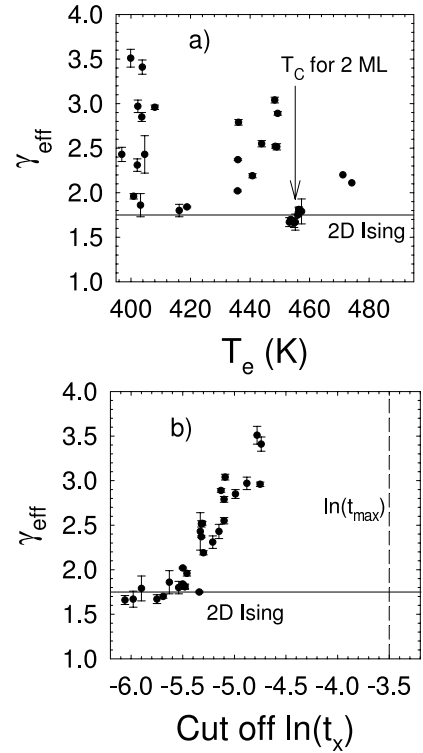}} 
\vspace{-.45in} 
\caption{A summary of experimental results from ref.\onlinecite{dunlavy} for $\gamma_{eff}$, determined from measurements of the in-plane easy axis magnetic susceptibility of Fe/W(110) films.  a) The value of the exponent is plotted against the peak temperature, $T_{e} \approx T_C$.  2D Ising values are obtained in a narrow temperature range corresponding to a Fe thickness of very nearly 2 ML.  b) The same data is plotted against $\ln t_x$, where $t_x$ is the fitted minimum limit of power law behavior on a ln-ln plot.  The maximum limit of power law behavior, due to signal levels, is indicated by the dashed line at $\ln t_{max}$.}
\end{figure}
study\cite{dunlavy} of the critical exponent, $\gamma$, of the magnetic susceptibility of Fe/W(110).  In this study of 25 films, in-plane easy axis ac magnetic susceptibility measurements have been fit objectively to determine four critical parameters simultaneously by the minimization of the variance from a simple power law in $\ln \chi$ vs. $\ln t$.  In addition to the intercept $\chi_0$ and slope $\gamma_{eff}$, both $T_C$ and a fitting cutoff $\ln t_x$ have been found.  $t_x$ defines the minimum of the range of reduced temperature over which the ln-ln plot is linear.  It defines how close to $T_C$ the power law extends before either extrinsic or intrinsic effects cause the susceptibility to deviate from a power law.  The maximum range of reduced temperature that is available, $t_{max}$, is determined by experimental signal-to-noise considerations.

The data from ref.[\onlinecite{dunlavy}] is replotted in two ways in fig. 6.
Part a) plots the fitted value of $\gamma_{eff}$ against the peak temperature, $T_{e}$, of the easy axis susceptibility (essentially the Curie temperature).  These data show that the 2D Ising value of $\gamma$ is recovered consistently and reproducibly only when $T_{e} \approx 455$K, which corresponds to the Curie temperature of 2.0 ML Fe films.   In fig. 6b, the same data is plotted against the cut-off $\ln t_x$.  $\ln t_{max}$ is indicated by the dashed line. Points further to the left on the plot represent susceptibility measurements that display power law scaling extending closer to $T_C$.  This plot shows two interesting correlations.  First, the Fe films where $\gamma_{eff}$ takes the 2D Ising value are not only very near 2 ML in thickness, they are also films which have the longest range of power law scaling.  Second, for films where $\gamma_{eff}$ is not the 2D Ising value, there is a systematic relationship such that the value of $\gamma_{eff}$ increases as the range of power law scaling is reduced.

For Fe/W(110) films that are very nearly 2 ML in thickness, comparison of fig. 6 and fig. 5b (using the right hand scale in $\ln t$) indicates that the inhomogeneities give no signal in the hard axis susceptibility  within the fitting region for $\gamma_{eff}$, and are therefore point-like in this range.  Nonetheless, the peak in the hard axis susceptibility at higher temperature shows that there is disorder, and the parameter $g$ is non-zero.   The fact that the fitted value of $\gamma_{eff}$ for the 2 ML films is that of the 2D Ising model indicates that $t_{cr} < t_x$, and the logarithmic corrections do not affect the fitting.  This establishes a limit of $g < -1/\ln(t_x) = 0.15$ for the 2 ML films.  This is midway between what numerical studies calibrate as strong disorder\cite{talapov} ($g=0.30$) and weak disorder\cite{roder} ($g=0.017$).  

For films that depart even 0.1 - 0.2 ML from a complete 2 ML, the situation is more complicated.  The peak in the hard axis susceptibility in fig. 5 moves to $\ln t =-4.5$, which is midway through the fitting region for $\gamma_{eff}$ in fig. 6.  In this circumstance, the inhomogeneities may not be point-like, and variations in the local Curie temperature may not be averaged out. At the same time, the increased disorder will increase $g$, moving $t_{cr}$ to the right.  If $\ln t_{cr} > \ln t_x$, the eq.(\ref{geff}) indicates that logarithimic corrections cause $\gamma_{eff} \approx 2.0$.  Since there is no independent method of determining $g$ for the films, this possibility cannot be tested.  By some combination of these two effects, the fitted value $\gamma_{eff}$ departs from the result for the 2D Ising model.

As the film thickness departs even further from 2 ML, the peak temperature of the hard axis susceptibility moves close to $t_x$, so that in the entire fitting region the inhomogeneities are not point-like.  The temperature range where they are point-like moves so close to $T_C$ that it is inaccessible to the experiment, and the question of logarithmic corrections does not arise.

The previous discussion indicates how the characteristic size of the defects causes the experimentally determined critical exponent to deviate from the value of the 2D Ising model, but it does not indicate why the deviation is always to a larger value, and why it is correlated to $\ln t_x$ in fig. 6b, but not $|\theta - \theta_0|$ in fig. 6a.  In ref.(\onlinecite{dunlavy}), the authors state that a systematic increase in $\gamma_{eff}$ as seen in the experiment is mimicked by  generating ``data'' numerically using a 2D Ising model with a Gaussian distribution of $T_C$ of width 2 to 5 K. As the range of $T_C$ is increased, so does $\gamma_{eff}$. This is entirely consistent with the present findings, where inhomogeneities create a mixed anisotropy. According to eq.(\ref{deltaTc}), a modest 5\% to 10\% change in the local anisotropy at the inhomogeneity can create the required range of Curie temperatures.  If the characteristic size of the inhomogeneities is small enough (very close to 2 ML), then the inhomogeneities are point-like in the critical region and the range of local values of $T_C$ is  not relevant.  However, moving away from 2 ML, the inhomogeneities are not point-like in the temperature region that is fitted for $\gamma_{eff}$, and the range of anisotropy values determines the range of $T_C$ and affects the fitted value of $\gamma_{eff}$.  Independent films of the same thickness can have a different distribution of anisotropies that produce a range of precisely determined, but different values of $\gamma_{eff}$, as is seen in fig. 6a.  However, regardless of thickness, the larger the range of $T_C$ in each film, the shorter will be the range in reduced temperature that will appear linear on a ln-ln scale.  Therefore, $\ln t_x$ is correlated with $\gamma_{eff}$ as in fig. 6b.

There are very few other published studies with which to compare these results.  Most critical studies of ultrathin magnetic films measure the exponent of the magnetization, $\beta$, in the ferromagnetic state below the Curie temperature.\cite{taroni}  In this case, the ferromagnetic correlation length is very large and the inhomogeneities are always point-like.  We are unaware of any other systematic studies of the critical susceptibility as a function of layer completion.  However, two studies of individual films are noteworthy. Jensen \emph{et al.}\cite{jensen} investigate a Co film on a vicinal Cu(1 1 17) surface, and find only a small hard axis susceptibility consistent with a homogeneous anisotropic 2D Heisenberg model.  The regularly-spaced lines of step edges of vicinal surfaces provide the dominant anisotropy, and do not form closed curves of characteristic size $d$.  We are unaware of measurements of critical exponents of films grown on this type of surface, but if theoretical calculations that consider the effect of infinite line defects in the 2D Ising model are applicable,\cite{mccoy,mccoy2,igloi} they may be nonuniversal.  

Finally, the results of Back \emph{et al.}\cite{back} seem to contradict our findings. They study a 1.7 ML film where the first monolayer is not annealed and the resulting $T_C \approx 340$K.  However, the film has an island structure, and, in a separate publication\cite{back2}, they find a broad in-plane hard axis susceptibility centered near $T_C$ when a relatively large field of 12 Oe is applied.  The fact that they measure the 2D Ising value of $\gamma$ may be a result of the use of a static method of measurement that can more closely approach $T_C$ than our ac method.\cite{dunlavy2}

\section{Conclusions}

Measurements of the magnetic susceptibility of Fe/W(110) films in the range of 1.6 to 2.4 ML Fe have revealed a small, hard axis response that is distinct from the much larger easy axis susceptibility associated with the phase transition to ferromagnetism.  The shape, size and peak temperature of the hard axis susceptibililty is not compatible with the 2D anisotropic Heisenberg model that is generally considered to describe a homogeneous, uniaxial, in-plane ferromagnet.  In addition, the dependence of the signal on the layer completion of the film as $|\theta - \theta_0|$, where $\theta_0$= 2.0 ML, is not consistent with a homogeneous system.  Rather, the hard axis signal indicates that magnetic inhomogeneties that remain relevant in the critical region create a system with mixed anisotropy. While the majority of the system can be described by the 2D anisotropic Heisenberg model (2D Ising in the critical region), a minority display a local easy axis behavior along the global hard axis.  Although the precise nature of the inhomogeneities  has not be identified, a likely candidate is the closed lines of monolayer step edges at the perimeter of the islands or pits associated with an incomplete layer.

Because the system has mixed anisotropy, the magnetic inhomogeneities affect the critical behavior of the film.  The variation of the peak amplitude and temperature of the hard axis susceptibility with $\sqrt{\theta - \theta_0}$ indicates that the inhomogeneities have a characterisitic size $d$.  Since the peak occurs when $\xi \approx d$, it marks a temperature range where the inhomogeneities are not point-like. In this case, the local variations in $T_C$ induced by the local changes in anisotropy are not averaged over, and lead to an increase in the value $\gamma_{eff}$ fitted to the susceptibility in a previous study.  Only when the Fe films are very close to a complete 2 ML is $d$ small enough that the inhomogeneities are point-like over the temperature range accessible to the experiments, and the fitted exponent has the 2D Ising value.  Although this makes it more difficult to access the asymptotic region experimentally, it offers an interesting opportunity to study the explicit magnetic response of defects and inhomogeneities. 

\section{Acknowledgements}

We are thankful for the continuing technical assistance of M. Kiela, for advice concerning experimental questions from N. Abu-Libdeh, and for insightful discussions with S.-S. Lee.  This work was supported by the Natural Sciences and Engineering Research Council of Canada.


\begin{thebibliography}{39}
\bibitem{remark2} The transverse Kerr effect causes changes in light intensity, rather than in polarization.  Unlike the longitudinal and polar Kerr effects, it does not exist to first order in film thickness, and is not allowed by symmetry for s-polarized light.
\end{thebibliography}
\end{document}